\documentclass[runningheads]{svmult}

\usepackage{makeidx}   
\usepackage{graphicx}  
\usepackage{subeqnar}  
\usepackage{multicol}  
\usepackage{physprbb}  
\makeindex             

\begin{document}
\title*{Obscured accreting black holes at high redshift}
\toctitle{Focusing of a Parallel Beam to Form a Point
\protect\newline in the Particle Deflection Plane}
%
%
\titlerunning{high--z obscured BH}
%
\author{Andrea Comastri\inst{1}
\and Fabrizio Fiore\inst{2} 
\and Cristian Vignali\inst{3}
\and Marcella Brusa\inst{4}
\and Francesca Civano\inst{3,1}}
%
\authorrunning{A. Comastri et al.}
%
%
\institute{INAF--Osservatorio Astronomico di Bologna, 
 via Ranzani 1, I--40127 Bologna, Italy
\and INAF--Osservatorio Astronomico di Roma,
 via Frascati 33, I--00040 Monteporzio (RM), Italy
\and Dipartimento di Astronomia, Universit\`a di Bologna, 
 via Ranzani 1, I--40127 Bologna, Italy
\and Max Planck Institut f\"ur Extraterrestrische Physik (MPE), 
      Giessenbachstrasse 1, D--85748 Garching,  Germany}

\maketitle              

\begin{abstract}
A significant fraction of the accreting black holes
powering  high redshift AGN are obscured by large columns of dust and gas.
For this reason, luminous 
type 2 quasars can be efficiently discovered 
combining hard X--ray and near--infrared observations.
We will briefly discuss the most recent results.

\end{abstract}

\section{Introduction}

The space density of luminous quasars in the distant Universe
represents a key observational ingredient to 
understand the formation and evolution of 
supermassive black holes (SMBHs).
The hard X--ray energy range (above 2 keV) is well suited
for this purpose as it provides an unbiased view
of the obscured accretion power which is responsible for the
majority of the SMBH energy output recorded in the X--ray background
(XRB) spectrum.
 
\begin{figure}[h]
\begin{center}
\includegraphics[width=0.6\textwidth]{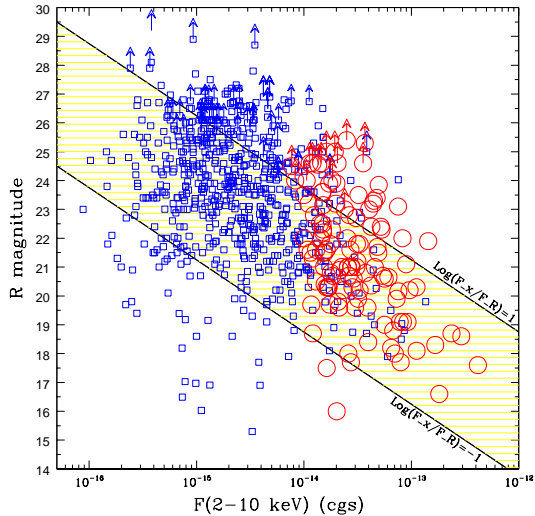}
\end{center}
\caption[]{The 2--10 keV flux vs. the R--band magnitude for 
X--ray selected sources in the Chandra (blue small squares) and 
HELLAS2XMM (red large open circles)}
\label{rhx}
\end{figure}

The capabilities of both {\it Chandra} and XMM--{\it Newton}
in performing sensitive X--ray surveys have been continuously 
exploited in the last four years.
As a result, large samples of X--ray sources spanning a wide range of 
fluxes (from a few 10$^{-13}$ down to about 10$^{-16}$ erg cm$^{-2}$ 
s$^{-1}$)  
are available for reliable statistical studies on the
extragalactic X--ray source population.
Despite extensive campaigns of spectroscopic follow--up
observations with the largest, ground--based telescopes, the X--ray source 
classification remains challenging. The most important reason 
is that the optical counterparts of many X--ray sources are 
too faint even for 8--10 m class telescopes (Fig.~\ref{rhx}).
The spectroscopic completeness of the deepest {\it Chandra} surveys
(CDFS and CDFN) is of the order of 50--70 \%. 
Deep multiband photometry is usually employed to estimate 
redshifts and, over restricted portions of the sky, 
a much higher completeness level could be achieved.

A detailed discussion on the optical identification breakdown 
is beyond the purposes of this paper and can be found elsewhere 
(\cite{barger03}, \cite{szokoly04}, \cite{fiore03}). 
The general picture emerging from the spectroscopic observations
can be summarized as follow:

\smallskip
\par\noindent
$\bullet$ Unobscured broad--line AGN are found up to redshift 5 
and constitute the dominant population at relatively bright X--ray
fluxes ($>$ 10$^{-14}$ erg cm$^{-2}$ s$^{-1}$) \cite{bauer04} 

\smallskip
\par\noindent 
$\bullet$ The space density of obscured AGN increases towards 
fainter fluxes. Clear signatures of X--ray and/or optical 
obscuration are typically
detected (though some mismatch between optical and X--ray absorbing
properties is present). The luminosity and redshift distribution
of obscured sources  
is skewed toward lower values and most of them are classified as Seyfert
galaxies ($L_X < 10^{44}$ erg s$^{-1}$ and $z < $ 1--1.5) rather than 
quasars.

\smallskip
\par\noindent 
$\bullet$ The discovery of 
relatively high X--ray luminosity AGN hosted in the nuclei of
optically normal galaxies. Whether they are heavily absorbed Seyfert 
galaxies, low--luminosity AGN whose emission is 
diluted by the host galaxy starlight 
or something more exotic is still subject of debate \cite{marcy03}.

\par
The most important findings emerging from the identification process
concern the redshift and absorption distribution 
of X--ray selected sources. The former is peaked at 
lower redshifts ($z \simeq$ 1) 
than expected on the basis of population synthesis models for the 
XRB \cite{gilli}. The latter seems to indicate a lack 
of obscured, luminous sources with respect to model predictions. 

On the one hand, the results described above 
favour a late formation of the XRB (i.e. \cite{steffen04}) 
and call for a major revision of the AGN synthesis models 
which are based on the unified scheme. 
On the other hand,  
it has been suggested \cite{treister} that selection effects against 
the identification of obscured AGN combined 
with the small solid angle covered by deep surveys 
prevent from a reliable comparison with model predictions.

\section{Large Area Surveys}

A sizable number of hard X--ray surveys
have been performed with both XMM--{\it Newton}
and {\it Chandra} over large area (of the order of a few
square degrees).
The ultimate goal of these efforts is to provide
a statistically robust estimate of the luminosity function and cosmological
evolution of obscured accreting black holes hopefully free 
from the selection effects described above.
Though differing in several details, the observing strategy of 
the various large--area surveys is designed 
to maximize the trade off between area and 
depth and thus X--ray satellite observing time
(for an updated list of recent X--ray surveys see \cite{brandt04}). 
Wide--area surveys efficiently 
target objects in the range of fluxes
around the knee of the source counts 
(log$F_X = - 14 \pm 1$ erg cm$^{-2}$ s$^{-1}$) 
and thus properly sample the sources which contribute mostly to the XRB.
An important benefit is that the average magnitude of the optical 
counterparts is relatively bright and the follow--up 
spectroscopic identification is 
within the capabilities of available telescopes.
On the other hand, the large number of X--ray sources 
discovered in many different non--contiguous pointings makes the 
follow--up identification extremely time consuming.

Targets with peculiar and/or extreme properties are therefore the subject 
of detailed follow--up multiwavelength observations.
The results of a vigorous program of spectroscopic observations 
with FORS@VLT \cite{fiore03}, deep near--infrared 
photometry with ISAAC@VLT \cite{migno04}, 
and X--ray spectroscopy \cite{perola04}, 
strongly indicate that sources with high 
X--ray to optical flux ratio (high $X/O$) in the 
{\tt HELLAS2XMM} survey \cite{baldi02} 
are obscured AGN at moderate to high redshifts.
The derived X--ray luminosities and the narrow optical lines 
of the brightest targets allow us to classify them as 
type 2 quasars at $z$=0.7--2. A similar range of
redshifts is implied using the R--K colour of the host galaxies
of fainter X--ray sources  \cite{migno04}. 

Thanks to the identification of about 50\% of the high $X/O$ sources in the
{\tt HELLAS2XMM} survey, it has been 
possible to discover a linear correlation between the 
$X/O$ ratio and 2--10 keV luminosity \cite{fiore03}, namely:
log $L_{2-10} = log f_X / f_{opt} + 43.05$.  
The correlation holds for optically obscured sources 
and has been tested and calibrated combining the 
optical and X--ray data of the {\tt HELLAS2XMM} survey
with well defined subsamples of identified sources 
in the deep {\it Chandra} fields at fluxes larger than 
3$\times 10^{-15}$ erg cm$^{-2}$ s$^{-1}$.

Though characterized by a not--negligible dispersion (about 0.4 dex), 
this relation can be used to compute X--ray luminosities 
and then redshifts from the observed $X/O$ ratio.
The accuracy in the redshift estimate (``X--photo--z'' \cite{fiore04})
is fairly good  $\sigma(\Delta z / (1+z)) \simeq$ 0.2.

\section{The high--redshift Universe} 

\begin{figure}[t]
\begin{center}
\includegraphics[width=0.6\textwidth]{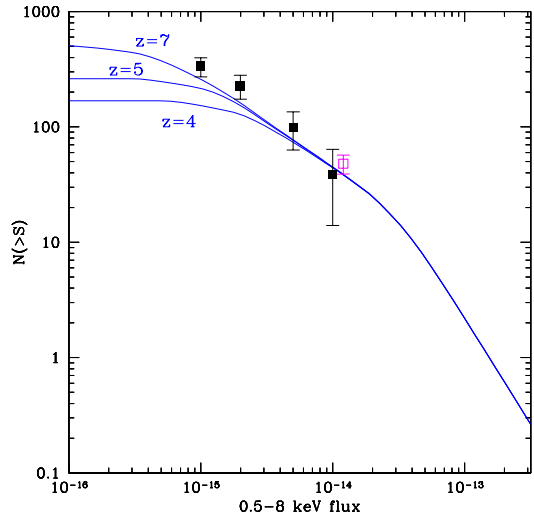}
\end{center}
\caption[]{The number counts of luminous [$L_X > 10^{44}$ erg s$^{-1}$] 
obscured ($N_H > 10^{22}$ cm$^{-2}$) quasars as estimated by
Perola et al. (2004, open square), and Padovani et al. (2004,  
filled squares). Continuous lines represent the expected number counts
from the XRB synthesis models as described in the text}
\label{qso2}
\end{figure}

The space density of high luminosity ($L_X > 10^{44}$ erg s$^{-1}$) 
obscured ($N_H >$ 10$^{22}$) 
type 2 quasars among optically faint {\it Chandra} deep 
fields sources has been estimated by Padovani et al. (2004) \cite{pado04}
using the ``X--photo--z'' technique.
In order to select absorbed sources, a simple criterium based 
on the Hardness Ratio [HR = (H--S)/(H+S) $>$ --0.2, where H and S is the
number of counts in the 2--8 keV and 0.5--2 keV bands, respectively] 
is adopted. About half (31/68) of the {\it Chandra} sources 
with an hard X--ray spectrum are candidate type 2 quasars.  
Their number counts,
computed assuming the $X/O$ vs $L_{2-10 keV}$ relation, are reported in 
Figure~\ref{qso2} along with the results obtained from 
the {\tt HELLAS2XMM} survey \cite{perola04}.

The observed surface densities are compared with the 
number counts of luminous ($L_{0.5-8 keV} > 10^{44}$ erg s$^{-1}$) 
obscured ($N_H > 10^{22}$ cm$^{-2}$) AGN predicted by
XRB synthesis models \cite{coma95}.
The evolution of the X--ray luminosity function is parameterized 
by a pure luminosity law [$L(z) \propto L(z=0) \times (1+z)^{2.6}$]
up to $z$ = 1.5 and constant up to a maximum redshift $z_{max}$.

Model predictions are shown in Figure~\ref{qso2} 
for three different values of the 
maximum redshift over which the integration of the XLF is performed.
At their face value the results indicate that 
in order to reproduce the number counts of type 2 quasars as predicted
by \cite{pado04}, a maximum redshift as high as $z_{max} \simeq$ 7 is 
required.

\begin{figure}[t]
\begin{center}
\includegraphics[width=.49\textwidth]{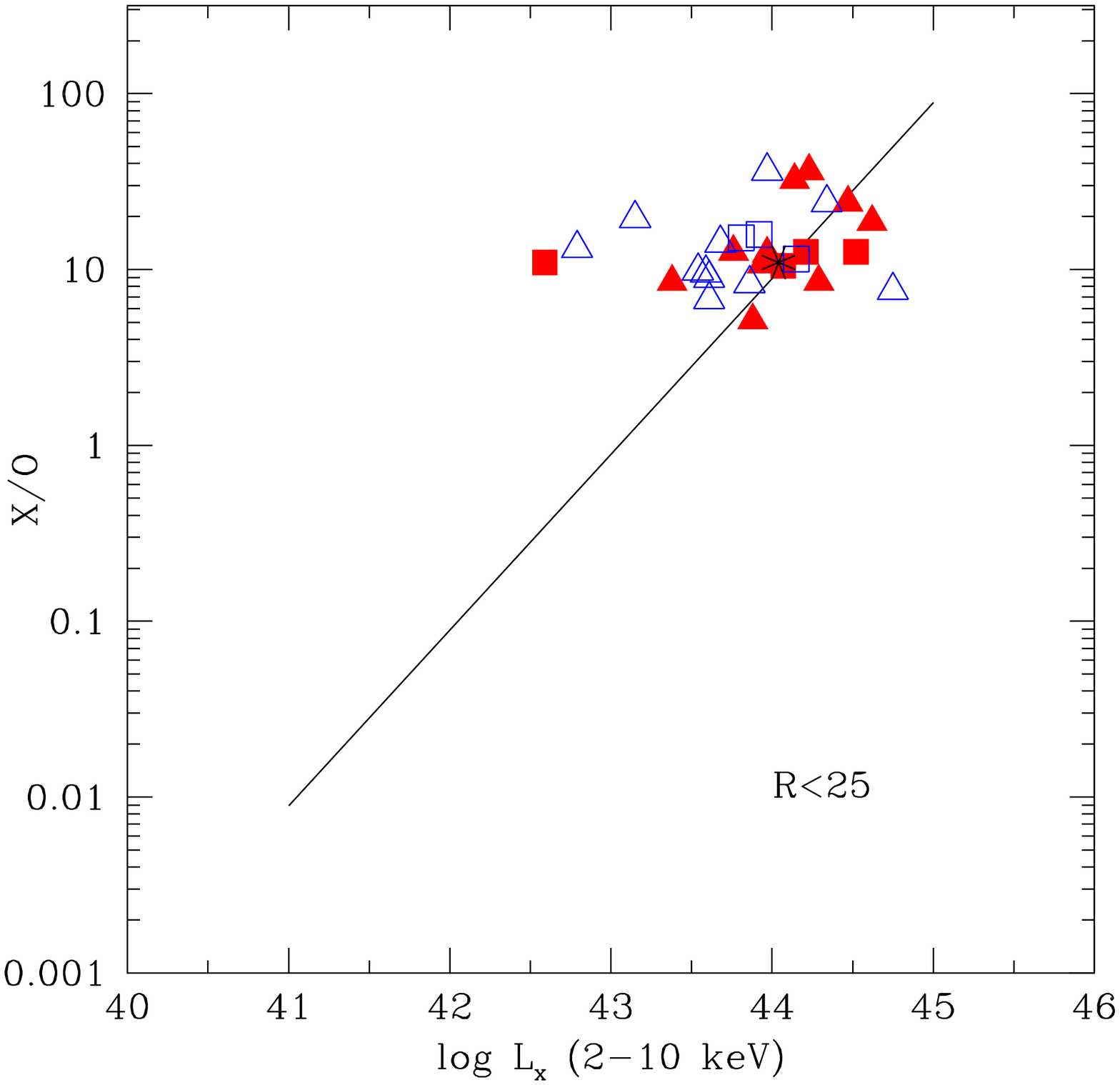}
\includegraphics[width=.49\textwidth]{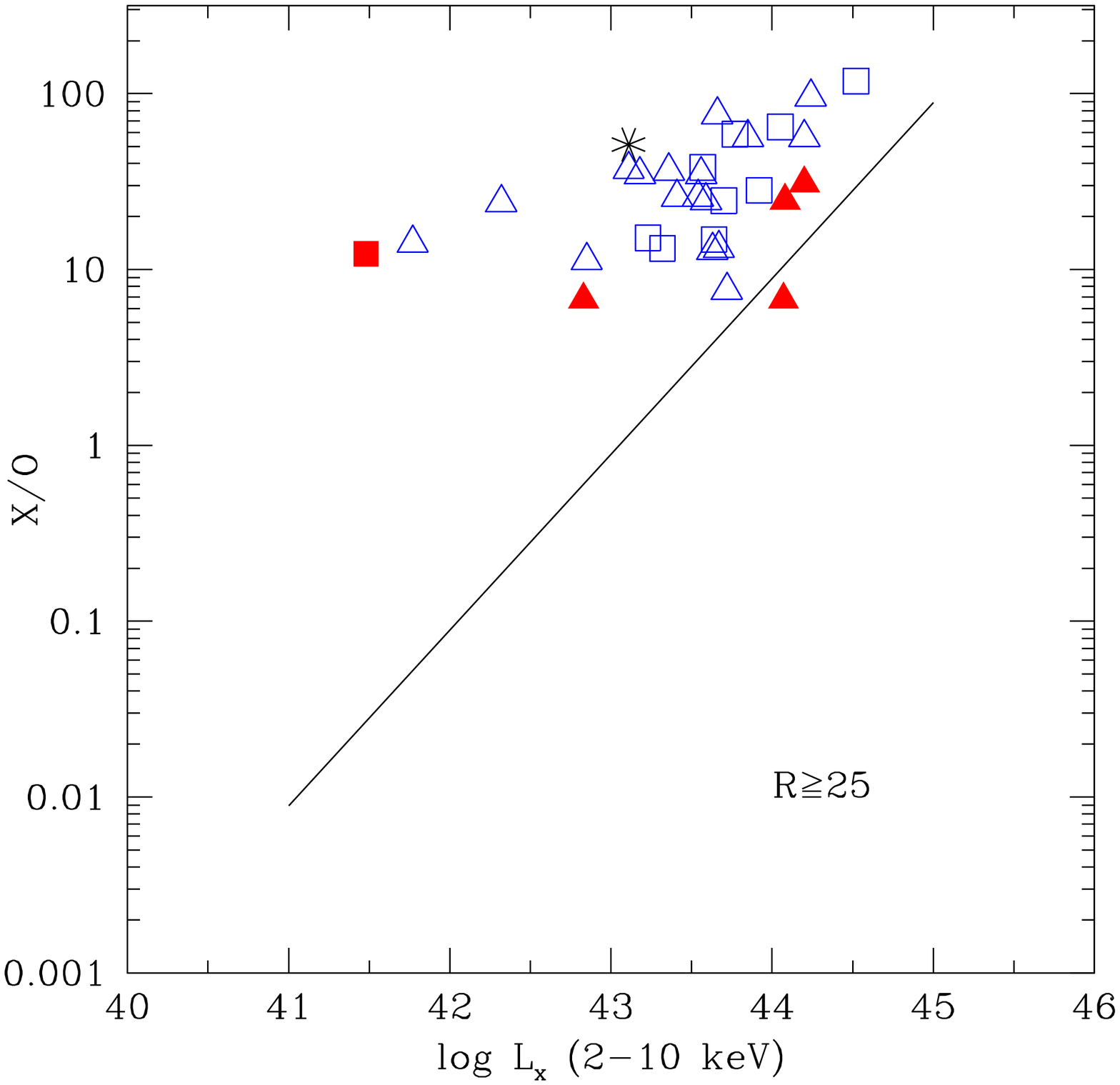}\end{center}
\caption[]{The X/O ratio as a function of 2--10 keV
absorption--corrected luminosity for a sample of hard X--ray selected
{\it Chandra} sources (CDFN = squares; CDFS = triangles; spectro--z = 
filled symbols; photo--z = open symbols; star = redshift obtained from the
iron line). {\it Left panel}: optically ``bright'' objects 
$R < $ 25, {\it Right panel}: optically faint objects $R > $ 25} 
\label{fra}
\end{figure}

The existence of a population of very high--redshift obscured
quasars hiding among the optically faint counterparts of X--ray 
sources has been put forward by \cite{anton}. Seven {\it Chandra} 
sources with robust detections in the X--ray band (25--90 counts) 
are not detected in deep multiband HST ACS observations.
Their extreme values of the $X/O$ ratio (EXO's) and in particular the lack of
detection in the $z_{850}$ ACS band are consistent with a redshift
above  6--7 such that their Ly$\alpha$ emission is redshifted out of 
the ACS bands. However making use of recent {\it Spitzer} 
observations in the IRAC 
filters (3.6--8.4 $\mu$m) and deep K$^{\prime}$ data, it has been 
suggested \cite{anton2} that their multiband photometric data 
are well fitted by early--type templates at redshifts 2--5, and in only
one case a redshift as high as 6 is supported by SED fitting. 

Given that only one spectroscopically confirmed quasar at 
$z>$ 5 has been so far discovered in the {\it Chandra} deep fields
\cite{barger03}, the possibility that even a few very high--z, 
presumably obscured, quasars could be hiding among high $X/O$ and EXO's  
would have important consequences for the AGN evolution.
However, before claiming that such a population has been indeed
revealed, additional and more robust observational evidences 
should be provided.

The identification of type 2 quasars using the ``X--photo--z''  technique 
is based on the assumption that the $X/O$ vs. L$_{2-10 keV}$ correlation 
can be extrapolated beyond the magnitudes over which it has been calibrated 
(R$<$ 24--25). In order to address this point, we \cite{francy05} 
have collected redshift measurements (mainly photo--z) 
for all the sources in the {\it Chandra} deep fields with $X/O > $ 10.
The results are reported in Figure~\ref{fra} where the sample
is divided according to the optical magnitude.
While relatively bright $R <$ 25 sources lie, though with a substantial 
scatter, on the relation, optically faint objects appear to have
lower luminosities than expected. Similar conclusions are reached 
by \cite{bauer04} (see their Figure 6). 
On the other hand, it is also important to note that
at faint optical magnitudes the probability to find by chanche a galaxy 
in the X--ray error box increases drammatically (i.e. up to about 0.25--0.30
for R=26 and an error circle radius of 2 arcsec, without considering
source clustering). Furthermore, photo--z estimates 
often involve the determination of source's 
magnitudes in images of very different quality (e.g. space versus ground
based telescopes) and, as a consequence, are affected by systematic errors.

Even if the $X/O$ ratio appears to be an efficient selection method to search
for high--z, obscured type 2 quasars \cite{marcy04}, 
a reliable estimate of their number and luminosity 
densities must await for medium--deep, large--area hard 
X--ray surveys allowing to sample, with an adequate 
statistic, the brightest sources. For a given $X/O$ their optical
counterparts are, on average, brighter and thus within the spectroscopic
capabilities of present telescopes. For extremely optically 
faint sources deep near--infrared photometry coupled with 
{\it Spitzer} observations would allow us to overcome the problem 
of chance coincidences.
Significant progress are foreseen once the rich multiwavelength 
database of the ongoing COSMOS (2 square degrees) 
and ELAIS--S1 (0.5 square degrees) surveys will be exploited.

%

\end{document}